\documentclass[10pt, a4paper]{article}
\usepackage{lrec2022} % this is the new LREC2022 Style
\usepackage{multibib}
\newcites{languageresource}{Language Resources}
\usepackage{graphicx}
\usepackage{tabularx}
\usepackage{soul}
\usepackage{titlesec}
\titleformat{\section}{\normalfont\large\bfseries\center}{\thesection.}{1em}{}
\titleformat{\subsection}{\normalfont\SmallTitleFont\bfseries\raggedright}{\thesubsection.}{1em}{}
\titleformat{\subsubsection}{\normalfont\normalsize\bfseries\raggedright}{\thesubsubsection.}{1em}{}
\renewcommand\thesection{\arabic{section}}
\renewcommand\thesubsection{\thesection.\arabic{subsection}}
\renewcommand\thesubsubsection{\thesubsection.\arabic{subsubsection}}
\usepackage{epstopdf}
\usepackage[utf8]{inputenc}

\usepackage{hyperref}
\usepackage{xstring}

\usepackage{color}

\usepackage{caption}
\usepackage{subcaption}
\usepackage{multirow,url,booktabs}
\newcommand{\scell}[2][c]{%
  \begin{tabular}[#1]{@{}c@{}}#2\end{tabular}}

\usepackage{tipa}

\usepackage[edges]{forest}

\definecolor{foldercolor}{RGB}{124,166,198}

\tikzset{pics/folder/.style={code={%
    \node[inner sep=0pt, minimum size=#1](-foldericon){};
    \node[folder style, inner sep=0pt, minimum width=0.3*#1, minimum height=0.6*#1, above right, xshift=0.05*#1] at (-foldericon.west){};
    \node[folder style, inner sep=0pt, minimum size=#1] at (-foldericon.center){};}
    },
    pics/folder/.default={20pt},
    folder style/.style={draw=foldercolor!80!black,top color=foldercolor!40,bottom color=foldercolor}
}

\forestset{is file/.style={edge path'/.expanded={%
    ([xshift=\forestregister{folder indent}]!u.parent anchor) |- (.child anchor)}, inner sep=0pt},
    this folder size/.style={edge path'/.expanded={%
    ([xshift=\forestregister{folder indent}]!u.parent anchor) |- (.child anchor) pic[solid]{folder=#1}}, inner xsep=0.75*#1},
    folder tree indent/.style={before computing xy={l=#1}},
    folder icons/.style={folder, this folder size=#1, folder tree indent=3*#1},
    folder icons/.default={12pt},
}
\title{KazakhTTS2: Extending the Open-Source Kazakh TTS Corpus\\With More Data, Speakers, and Topics}

\name{Saida Mussakhojayeva, Yerbolat Khassanov, Huseyin Atakan Varol}

\address{Institute of Smart Systems and Artificial Intelligence (ISSAI),\\Nazarbayev University, Nur-Sultan, Kazakhstan\\
         \{saida.mussakhojayeva, yerbolat.khassanov, ahvarol\}@nu.edu.kz\\}

\abstract{
We present an expanded version of our previously released Kazakh text-to-speech (KazakhTTS) synthesis corpus.
In the new KazakhTTS2 corpus, the overall size has increased from 93 hours to 271 hours, the number of speakers has risen from two to five (three females and two males), and the topic coverage has been diversified with the help of new sources, including a book and Wikipedia articles.
This corpus is necessary for building high-quality TTS systems for Kazakh, a Central Asian agglutinative language from the Turkic family, which presents several linguistic challenges.
We describe the corpus construction process and provide the details of the training and evaluation procedures for the TTS system.
Our experimental results indicate that the constructed corpus is sufficient to build robust TTS models for real-world applications, with a subjective mean opinion score ranging from 3.6 to 4.2 for all the five speakers.
We believe that our corpus will facilitate speech and language research for Kazakh and other Turkic languages, which are widely considered to be low-resource due to the limited availability of free linguistic data.
The constructed corpus, code, and pretrained models are publicly available in our GitHub repository.
\\\newline  }

\begin{document}

\maketitleabstract

%##############################################################################################################################################################################################################
\section{Introduction}
Text-to-speech (TTS), also known as speech synthesis, is the automatic process of converting written text into speech~\cite{taylor2009text}, which has wide application potential and a substantial social impact, including digital assistants, improved accessibility for people with reading disabilities, speech and vision impairments, to name a few.
For visually impaired people, in particular, it enables voice-controlled access to Internet-of-things devices, on-demand access to books and websites, and access to other vocalized assistive technologies.
In turn, these enhance the overall quality of life, consumption of information, and access to knowledge.
In addition, TTS can complement other important language and vision technologies, such as speech recognition~\cite{tjandra2017listening}, speech-to-speech translation~\cite{wahlster2013verbmobil}, face-to-face translation~\cite{DBLP:conf/mm/RMPJNJ19}, and visual-to-sound~\cite{Zhou_2018_CVPR}.
Considering the aforementioned benefits, TTS is undoubtedly an essential speech processing technology for any language. 

In recent years, TTS research has progressed remarkably thanks to neural network-based architectures~\cite{tan2021survey}, regularly organized challenges~\cite{DBLP:conf/interspeech/BlackT05,DBLP:conf/interspeech/DunbarAKBBCMDOB19}, and open-source datasets~\cite{ljspeech17,DBLP:conf/interspeech/ZenDCZWJCW19,DBLP:journals/corr/abs-2010-11567}.
Especially, impressive results have been achieved for commercially viable languages, such as English and Mandarin.
However, there is still a lack of research into the development of TTS technologies for low-resource languages.
To address this problem in regard to Kazakh, \newcite{mussakhojayeva21_interspeech} have recently developed the first open-source Kazakh text-to-speech (KazakhTTS) corpus, which contains 93 hours of manually transcribed audio from two professional speakers (one female and one male) reading news articles.
The developed corpus has generated substantial interest and has been downloaded over 200 times in less than a year by academia and industry, both from local and global organizations.
This demonstrates high demand for open-source and high-quality transcribed speech data in the Kazakh language.

Motivated by this, in this paper, we present a new version of the KazakhTTS corpus called KazakhTTS2, which adds more data, speakers, and topics to our corpus.
Specifically, we have increased the data size from 93 hours to 271 hours.
We have added three new professional speakers (two females and one male), with over 25 hours of transcribed data for each speaker.
In addition to news, we have diversified the topic coverage with a book and Wikipedia articles.
%Like the first version, KazakhTTS2 is freely available\footnote{\url{https://github.com/IS2AI/Kazakh_TTS}\label{ft:github}} to both academic researchers and industry practitioners under the CC BY 4.0 license\footnote{\url{https://creativecommons.org/licenses/by/4.0/}}.
Like the first version, KazakhTTS2 is freely available to both academic researchers and industry practitioners in our GitHub repository\footnote{\url{https://github.com/IS2AI/Kazakh_TTS}\label{ft:github}}.

To validate the KazakhTTS2 corpus, we built a state-of-the-art TTS system based on the Tacotron~2~\cite{DBLP:conf/icassp/ShenPWSJYCZWRSA18} architecture.
The constructed TTS system was evaluated using the subjective mean opinion score (MOS) measure.
%The obtained MOSs for all the speakers were above 4.0, which indicates the utility of the KazakhTTS2 corpus for building robust TTS systems suitable for real-world applications.
The obtained MOSs for all the speakers ranged from 3.6 to 4.2, which indicates the utility of the KazakhTTS2 corpus for building robust TTS systems suitable for real-world applications.
We believe that our corpus will further facilitate the rapid development of TTS systems in the Kazakh language and thus serve as an enabler for the wide range of applications mentioned above.
%We also hope that this work encourages subsequent efforts in this area for other similar languages from the Turkic family.
We also believe that this work will encourage subsequent efforts in this area to address some of the practical issues that arise when training TTS systems for the Kazakh language. % which is an agglutinative and employs vowel harmony.
Additionally, our corpus can be employed to bootstrap speech technologies for other similar languages from the Turkic family, for example, by means of cross-lingual transfer learning~\cite{DBLP:conf/interspeech/ChenTYL19} and self-supervised pretraining~\cite{NEURIPS2020_92d1e1eb}.

To sum up, our main contributions are:
\vspace{-0.2cm}
\begin{itemize}
    \item We developed a text-to-speech synthesis corpus for the Kazakh language containing five speakers (three females and two males) comprising 271 hours of carefully transcribed data from various sources (news, book, and Wikipedia).\vspace{-0.2cm}
    \item We validated the efficacy of the corpus, by training state-of-the-art neural TTS models, which achieved a sufficient subjective MOS for most practical applications.\vspace{-0.2cm}
    \item The KazakhTTS2 corpus, code, and pretrained models were made publicly available\textsuperscript{\ref{ft:github}} for both commercial and academic use.
    \vspace{-0.2cm}
    %\item The KazakhTTS2 corpus was made publicly available\textsuperscript{\ref{ft:github}} under the CC BY 4.0 license with no restrictions on commercial or academic use.
    %To enable experiment reproducibility, we also shared our training recipes and pretrained models.
    %\vspace{-0.2cm}
\end{itemize}

The rest of this paper is organized as follows:
Section~\ref{sec:rel} reviews the work on Kazakh language corpus creation.
In Section~\ref{sec:corpus}, we briefly summarize the previous release of the corpus and explain the changes made in KazakhTTS2, including the corpus structure and statistics.
The experimental setup and evaluation results are described in Section~\ref{sec:exp}.
Section~\ref{sec:discuss} discusses the challenges of Kazakh speech synthesis and future research directions.
Section~\ref{sec:conclude} concludes this work.

%##############################################################################################################################################################################################################
\section{Related Work}\label{sec:rel}
Despite its under-resourced status, Kazakh language research is emerging as an evolving field with an increasing number of recently released open-source corpora.
For example, \newcite{khassanov-etal-2021-crowdsourced} developed the first large-scale publicly available corpus for automatic speech recognition (ASR).
The corpus was collected by means of crowdsourcing, with over 2,000 people contributing around 330 hours of audio recordings.
Similarly, \newcite{yeshpanov2021kaznerd} developed an open-source Kazakh named entity recognition dataset consisting of over 100,000 sentences annotated for 25 entity classes.
Linguistic corpora development has also been observed in neighboring countries with languages similar to Kazakh, such as Uzbek~\cite{DBLP:conf/specom/MusaevMKKOV21}.
Additionally, there are other large-scale projects aimed at collecting open-source corpora for various languages, including Kazakh, such as Common Voice~\cite{DBLP:conf/lrec/ArdilaBDKMHMSTW20}.
However, all these datasets are unsuitable for building robust Kazakh TTS systems, which require a large number of high-quality audio recordings of a single speaker.

The first attempt to collect a large-scale open-source TTS dataset for the Kazakh language was made by~\newcite{mussakhojayeva21_interspeech}.
The collected dataset was called KazakhTTS and consisted of 93 hours of carefully transcribed audio from two professional speakers.
Specifically, the speakers were assigned to read local news articles.
The recorded articles were manually segmented into sentences and then aligned with the corresponding text with the help of native Kazakh transcribers.
The TTS systems developed using KazakhTTS achieved an MOS of above 4.0, demonstrating the high quality of the collected data. 
This work further extends the KazakhTTS corpus, as described in the following sections. 

The other existing corpora dedicated to Kazakh TTS are either proprietary or have been collected by leveraging unsupervised and semi-supervised approaches.
For example, \newcite{DBLP:conf/icassp/Black19} extracted readings of the Bible in hundreds of languages, including Kazakh.
The extracted recordings were automatically segmented and aligned with the corresponding text.
Although a TTS system built using this corpus is sufficient to deploy in some use-cases, its overall quality is unsatisfactory for most real-world applications.
Specifically, in the evaluation experiments, the Kazakh TTS system achieved a mel-cepstral distortion score of more than 6, which is considered low quality\footnote{\url{http://festvox.org/cmu_wilderness/index.html}}.
In another work, \newcite{DBLP:conf/specom/KhomitsevichMTR15} developed a Kazakh TTS system using a female voice.
However, the authors did not provide any information on their corpus, such as its size, how the recordings were acquired, and how to download it.
Additionally, the authors did not describe the evaluation procedures performed to assess the developed TTS system.
%In another work, \newcite{DBLP:conf/specom/KhomitsevichMTR15} developed a Kazakh TTS system using female voice recordings covering around 5.6 hours, which may not be sufficient to build robust Kazakh TTS systems.
%The authors did not provide information on the evaluation procedures performed to assess the developed TTS system or how the recordings were acquired.
%Additionally, the authors did not mention the ways to obtain their data.

%##############################################################################################################################################################################################################
\section{KazakhTTS2 Corpus}\label{sec:corpus}
In this section, we describe the curation procedures for the KazakhTTS2 corpus.
The KazakhTTS2 corpus collection was approved by the Institutional Research Ethics Committee of Nazarbayev University.
We first briefly summarize the previous version of the corpus (i.e., KazakhTTS) and then systematically explain the changes made to extend it.

\subsection{KazakhTTS}
KazakhTTS is the first version of our corpus, which contains around 93 hours of transcribed audio consisting of over 42,000 sentences.
The audio was recorded by two professional speakers, both of whom had had over ten years of narration experience in local television and radio stations.
The speakers were assigned to read news articles covering various topics, such as sports, business, politics, and so on.
The recorded audio was manually segmented into sentences, with defective segments (e.g., mispronunciation and external noise) filtered out.
The correspondence between audio and text was verified by native Kazakh transcribers.
The statistics for the first and second versions of the corpus are provided in Table~\ref{tab:compare}.

\begin{table}[th]
    \begin{center}
        %\small
        \renewcommand\arraystretch{1.1}
        \setlength{\tabcolsep}{3.0mm}
        \begin{tabularx}{\columnwidth}{l|cc}
            \toprule
            \textbf{Category}                   & \textbf{KazakhTTS}    & \textbf{KazakhTTS2}   \\
            \midrule
            \# Speakers                         & 2                     & 5 \\
            \# Segments                         & 42,082                & 136,196 \\
            \# Tokens                           & 565.6k                & 1.7M\\
            \# Unique tokens                    & 54.9k                 & 107.9k\\
            Duration                            & 93.2 h                & 271.7 h \\
            \bottomrule
        \end{tabularx}
        \caption{The comparison of statistics for the KazakhTTS and KazakhTTS2 corpora}
        \label{tab:compare}
    \end{center}
\end{table}

\subsection{Text Collection}
%While the previous corpus version was created by narrating the articles extracted from a single news website, in the new version, we collected additional articles from four local news websites.
We began by collecting additional news articles from four local news websites.
To further broaden the topic coverage, we added a book from the public domain and Wikipedia articles.
From Wikipedia, we extracted articles on science, computer technology, countries, and history.
All the articles were manually extracted to eliminate defects peculiar to web crawlers and saved in the DOC format for the professional speakers' convenience (i.e., font size, line spacing, and typeface could be adjusted to the preferences of the speakers).
In total, over 2,500 additional news articles, one book, and 159 Wikipedia articles were extracted.

\subsection{Recording Process}
To narrate the collected text, we auditioned several candidates and, as a result, hired three professional speakers (two females and one male).
Each speaker participated voluntarily and was informed of the protocols for data collection and use through an informed consent form.
All the hired speakers were tasked to read news articles only.
In addition, we rehired the male speaker (speaker M1) from the previous corpus creation process, because of his extensive experience in narrating documentaries. 
He was subsequently tasked to read the book and Wikipedia articles.
%A data release consent form was signed by each speaker before the recording session.
The speaker specifications, including gender, age, professional experience as a narrator, and recording device information, are provided in Table~\ref{tab:spk}.
Speakers F1 and M1 were part of KazakhTTS, whereas F2, F3, and M2 are newly hired speakers.

Due to the COVID-19 pandemic, we could not invite the speakers to our laboratory for data collection.
%Therefore, the speakers were allowed to record audio in their own studio at home setup for working from home purposes.
%Therefore, the speakers were allowed to record audio in their homemade studio set up for working from home purposes.
Therefore, the speakers were allowed to record audio in their makeshift studios that they had set up to work from home.
%Thus, F1 speaker recorded audios at her office studio, whereas other speakers recorded audios at home.
%Thus, all speakers recorded audios at home.
The speakers were instructed to read the texts in a quiet indoor environment with neutral tone and pace.
They were also asked to follow orthoepic rules, to maintain a constant distance between the microphone and lips, to pause at commas, and to intonate sentences ending with a question mark appropriately.
In total, each newly hired speaker read around 1,400 news articles, and Speaker M1 read one book entitled \textit{Abai Zholy} (The Path of Abai) and 159 Wikipedia articles.

%see Figure 7 of Burmese Speech Corpus, Finite-State Text Normalization and Pronunciation Grammars with an Application to Text-to-Speech

\begin{table}[h]
    \begin{center}
        \small
        \renewcommand\arraystretch{1.1}
        \setlength{\tabcolsep}{1.2mm}
        \begin{tabularx}{\columnwidth}{c|c|c|c|c}
            \toprule
            \textbf{Speaker ID} & \textbf{Gender}   & \textbf{Age}  & \textbf{\scell{Work\\experience}} & \textbf{\scell{Recording\\device}} \\
            \midrule
            \ \ F1$^*$          & Female            & 44            & 14 years                          & AKG P120 \\ % Raushan, Home - HyperX SoloCast, Office - AKG P120
            F2                  & Female            & 39            & 15 years                          & RØDE \\ % Assel, 
            F3                  & Female            & 52            & 25 years                          & Behringer C-1 \\ % Gulzhanat, 
            \ \ M1$^*$          & Male              & 46            & 12 years                          & Tascam DR-40 \\ % Islamkhan
            M2                  & Male              & 33            & 11 years                          & Mi \\ % Duman
            \bottomrule
        \end{tabularx}
        
        \vspace{0.1cm}
        {\raggedright \textit{Note.} Speakers F1 and M1 are from KazakhTTS.\par}
        \caption{The KazakhTTS2 speaker information}\label{tab:spk}
    \end{center}
\end{table}

\subsection{Segmentation and Alignment}
For audio segmentation and audio-to-text alignment, we employed the same approach as in the KazakhTTS corpus construction.
%We hired five native Kazakh transcribers, who manually segmented the recordings into sentence-level chunks and aligned them with the corresponding text using the Praat toolkit~\cite{boersma2001praat}.
We hired five native Kazakh transcribers with different backgrounds and thorough knowledge of Kazakh grammar rules.
The transcribers manually segmented the recordings into sentence-level chunks and aligned them with the corresponding text using the Praat toolkit~\cite{boersma2001praat}.
All the texts were represented using a Cyrillic script consisting of 42 letters\footnote{Note that at the time of writing, the Cyrillic alphabet is the official alphabet used for the Kazakh language, though the transition process to the Latin alphabet has already begun.} and other punctuation marks, such as period, comma, hyphen, question mark, and exclamation mark.
The transcribers were instructed to remove segments with mispronunciation and background noise, to trim long pauses at the beginning and end of segments, and to convert numbers and special characters (e.g., `\%', `\$', `+', etc) into the written form.
To ensure the uniform quality of work among the transcribers, we assigned a linguist to randomly check the completed tasks and to organize regular ``go through errors" sessions.

To ensure the correctness of the audio-to-text alignment process, the segmented recordings were inspected using our internal ASR system trained on the KSC dataset~\cite{khassanov-etal-2021-crowdsourced}.
Specifically, the ASR system was used to generate segment transcriptions, which were then compared to the corresponding manually annotated transcripts.
Segments with a high character error rate (CER) were regarded as incorrectly transcribed, and therefore rechecked by the linguist.
%Lastly, we filtered out all segments containing international words written using a non-Cyrillic alphabet, because speakers usually don't know how to correctly pronounce such words.

\subsection{Corpus Structure and Statistics}

\begin{figure} 
    \centering 
    %\tiny
    \scriptsize
    \begin{forest}
        for tree={
            font=\sffamily,
            grow'=0,
            inner sep=1pt,
            folder indent=0.0em,
            folder icons,
            edge=densely dotted
        }
        [KazakhTTS2, this folder size=-30pt
            [F1, this folder size=8pt
                [Transcripts, this folder size=8pt
                    [source\_docID\_uttID.txt, is file]]
                [Audio, this folder size=8pt
                    [source\_docID\_uttID.wav, is file]]]
            [F2, this folder size=8pt
                [Transcripts, this folder size=8pt
                    [source\_docID\_uttID.txt, is file]]
                [Audio, this folder size=8pt
                    [source\_docID\_uttID.wav, is file]]]
            [F3, this folder size=8pt
                [Transcripts, this folder size=8pt
                    [source\_docID\_uttID.txt, is file]]
                [Audio, this folder size=8pt
                    [source\_docID\_uttID.wav, is file]]]
            [M1, this folder size=8pt
                [News, this folder size=8pt
                    [Transcripts, this folder size=8pt
                        [source\_docID\_uttID.txt, is file]]
                    [Audio, this folder size=8pt
                        [source\_docID\_uttID.wav, is file]]]
                [Wiki, this folder size=8pt
                    [Transcripts, this folder size=8pt
                        [source\_docID\_uttID.txt, is file]]
                    [Audio, this folder size=8pt
                        [source\_docID\_uttID.wav, is file]]]
                [Book, this folder size=8pt
                    [Transcripts, this folder size=8pt
                        [source\_docID\_uttID.txt, is file]]
                    [Audio, this folder size=8pt
                        [source\_docID\_uttID.wav, is file]]]]
            [M2, this folder size=8pt
                [Transcripts, this folder size=8pt
                    [source\_docID\_uttID.txt, is file]]
                [Audio, this folder size=8pt
                    [source\_docID\_uttID.wav, is file]]]
            [speaker\_metadata.txt, is file]
        ]
    \end{forest}
    \caption{The file structure of KazakhTTS2}\label{fig:file_struc}
\end{figure}
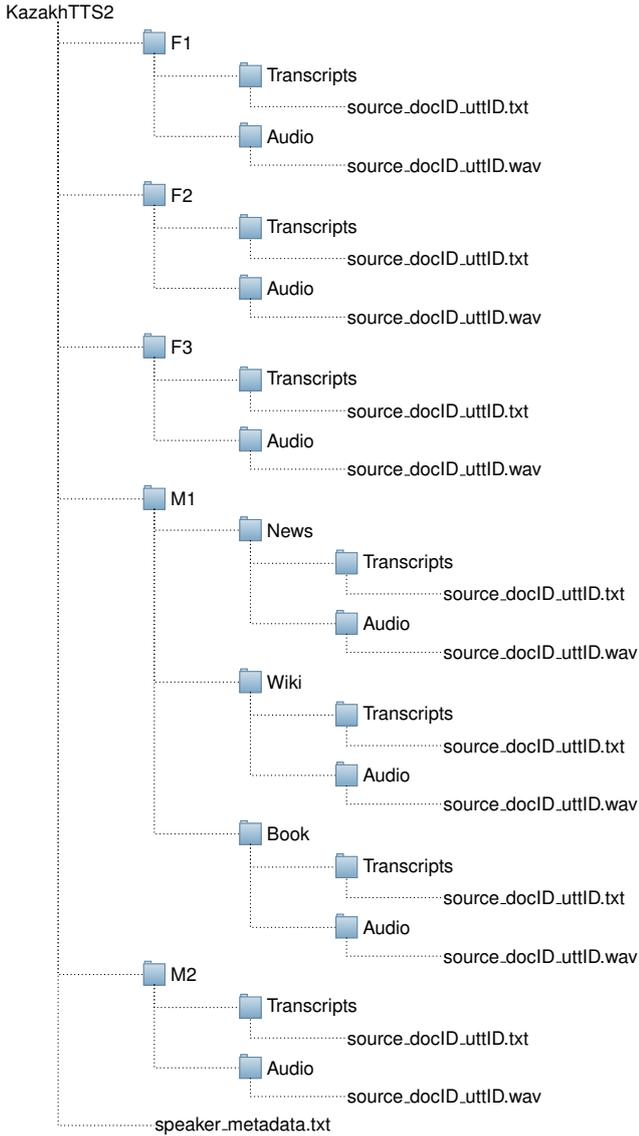

The file structure of the KazakhTTS2 corpus is shown in Figure~\ref{fig:file_struc}.
Collections of audio recordings and the corresponding transcriptions are stored in a separate folder for each speaker.
Additionally, for Speaker M1, we split the data from different sources into separate folders (i.e., News, Wiki, and Book).
All audio recordings were downsampled to 22.05 kHz and stored at 16 bits per sample in the WAV format.
All transcripts are stored as TXT files in the UTF-8 encoding.
The audio and the corresponding transcript filenames are identical except for the extension.
The name of each file consists of the source name, document ID, and utterance ID (i.e., \textit{source\_docID\_uttID}).
Speaker information, including gender, age, professional experience, and recording device, is provided in the \textit{speaker\_metadata.txt} file.

\begin{figure*}
    \centering
    \begin{subfigure}[b]{0.195\textwidth}
        \centering
        \includegraphics[width=\textwidth,trim={0.6cm 0.5cm 0.5cm 0.3cm},clip=true]{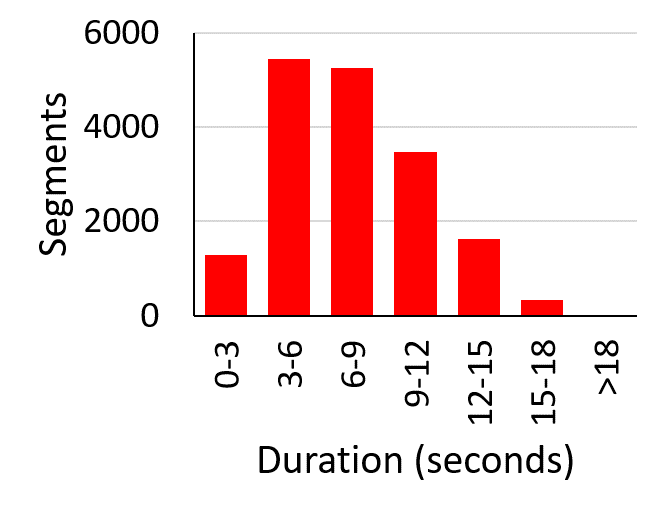}
        \vspace{-0.5cm}
        \caption{F1}
        \vspace{-0.2cm}
        \label{fig:y equals x}
    \end{subfigure}
    \hfill
    \begin{subfigure}[b]{0.195\textwidth}
        \centering
        \includegraphics[width=\textwidth,trim={0.6cm 0.5cm 0.5cm 0.3cm},clip=true]{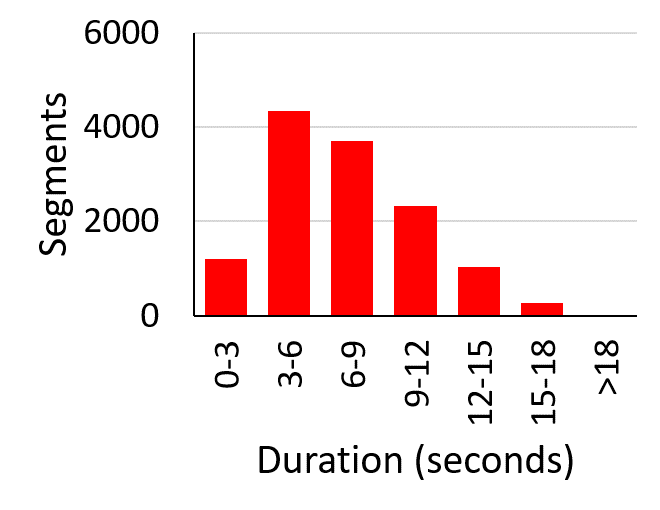}
        \vspace{-0.5cm}
        \caption{F2}
        \vspace{-0.2cm}
        \label{fig:five over x}
    \end{subfigure}
    \hfill
    \begin{subfigure}[b]{0.195\textwidth}
        \centering
        \includegraphics[width=\textwidth,trim={0.6cm 0.5cm 0.5cm 0.3cm},clip=true]{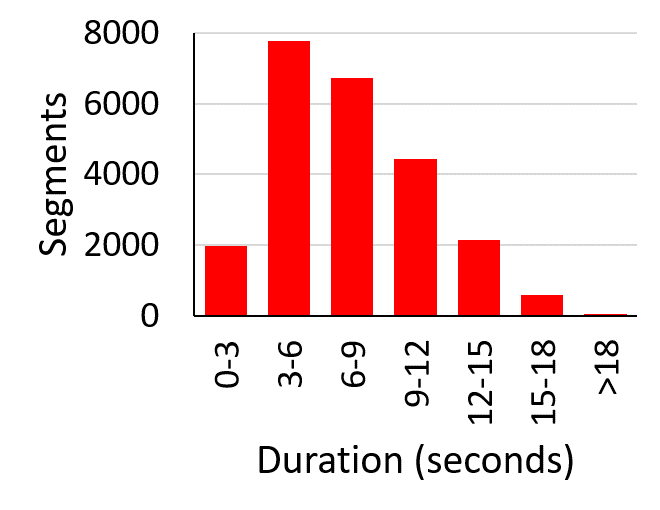}
        \vspace{-0.5cm}
        \caption{F3}
        \vspace{-0.2cm}
        \label{fig:five over x}
    \end{subfigure}
    \hfill
    \begin{subfigure}[b]{0.195\textwidth}
        \centering
        \includegraphics[width=\textwidth,trim={0.6cm 0.5cm 0.5cm 0.3cm},clip=true]{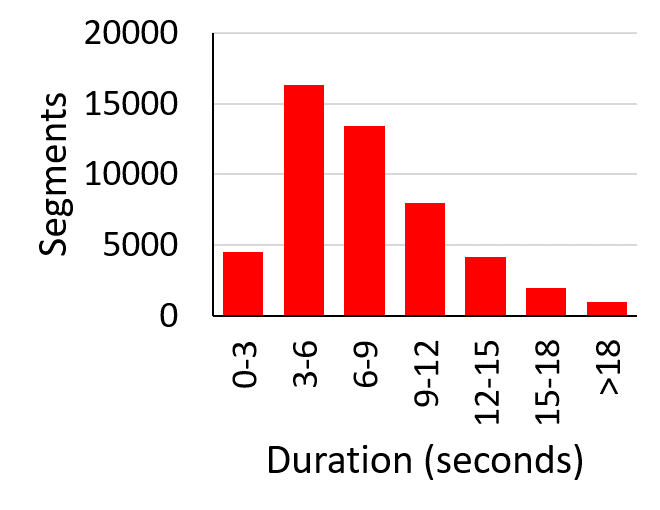}
        \vspace{-0.5cm}
        \caption{M1}
        \vspace{-0.2cm}
        \label{fig:five over x}
    \end{subfigure}
    \hfill
    \begin{subfigure}[b]{0.195\textwidth}
        \centering
        \includegraphics[width=\textwidth,trim={0.6cm 0.5cm 0.5cm 0.3cm},clip=true]{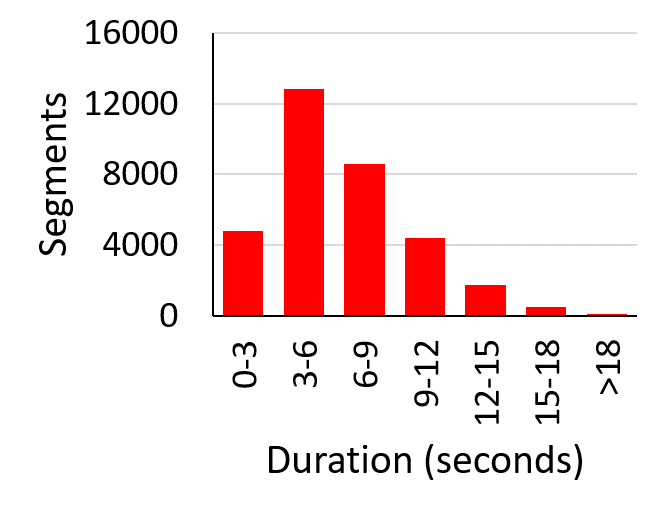}
        \vspace{-0.5cm}
        \caption{M2}
        \vspace{-0.2cm}
        \label{fig:five over x}
    \end{subfigure}
    \hfill

    \begin{subfigure}[b]{0.195\textwidth}
        \centering
        \includegraphics[width=\textwidth,trim={0.6cm 0.5cm 0.5cm 0.3cm},clip=true]{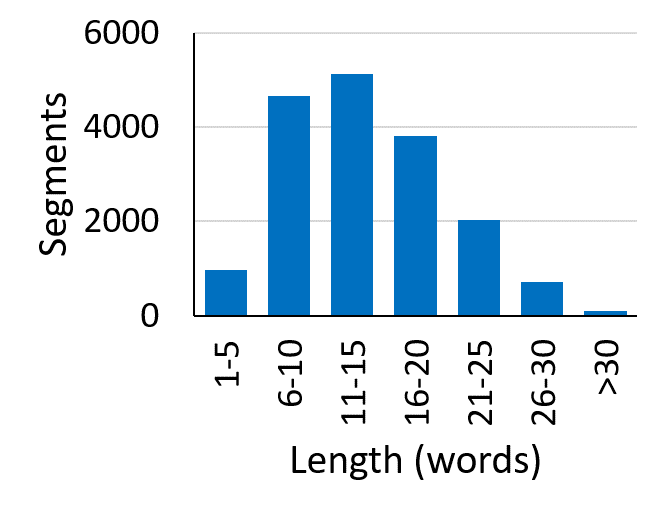}
        \vspace{-0.5cm}
        \caption{F1}
        %\vspace{0.2cm}
        \label{fig:three sin x}
    \end{subfigure}
    \hfill
    \begin{subfigure}[b]{0.195\textwidth}
        \centering
        \includegraphics[width=\textwidth,trim={0.6cm 0.5cm 0.5cm 0.3cm},clip=true]{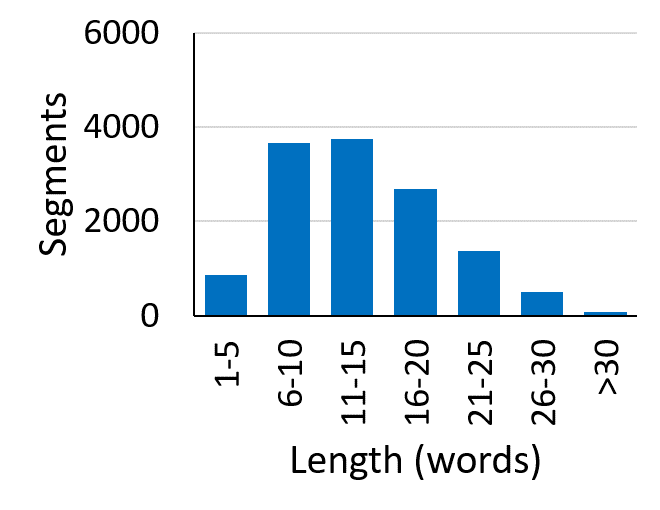}
        \vspace{-0.5cm}
        \caption{F2}
        %\vspace{0.2cm}
        \label{fig:five over x}
    \end{subfigure}
    \hfill
    \begin{subfigure}[b]{0.195\textwidth}
        \centering
        \includegraphics[width=\textwidth,trim={0.6cm 0.5cm 0.5cm 0.3cm},clip=true]{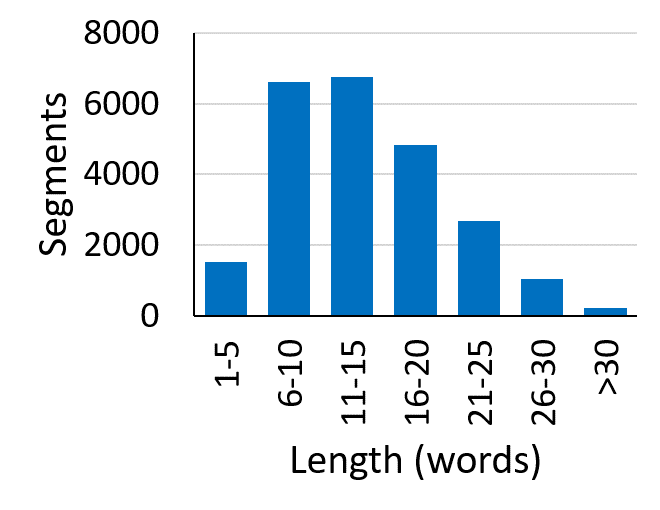}
        \vspace{-0.5cm}
        \caption{F3}
        %\vspace{0.2cm}
        \label{fig:five over x}
    \end{subfigure}
    \hfill
    \begin{subfigure}[b]{0.195\textwidth}
        \centering
        \includegraphics[width=\textwidth,trim={0.6cm 0.5cm 0.5cm 0.3cm},clip=true]{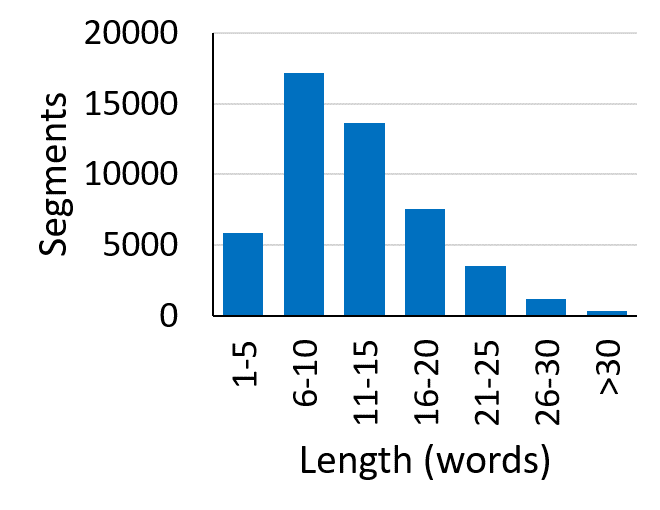}
        \vspace{-0.5cm}
        \caption{M1}
        %\vspace{0.2cm}
        \label{fig:five over x}
    \end{subfigure}
    \hfill
    \begin{subfigure}[b]{0.195\textwidth}
        \centering
        \includegraphics[width=\textwidth,trim={0.6cm 0.5cm 0.5cm 0.3cm},clip=true]{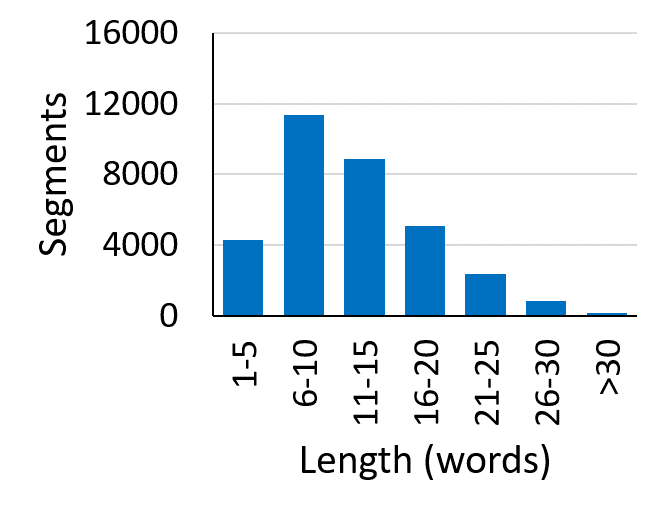}
        \vspace{-0.5cm}
        \caption{M2}
        %\vspace{0.2cm}
        \label{fig:five over x}
    \end{subfigure}
    \caption{Segment duration (a, b, c, d, e) and length (f, g, h, i, j) distributions for each speaker of KazakhTTS2}
    \label{fig:dist}
\end{figure*}

The statistics for the KazakhTTS2 corpus are given in Table~\ref{tab:data}.
The overall corpus size is around 271 hours, with each speaker having at least 25 hours of transcribed audio.
The total number of sentences is around 136 thousand, and the total number of tokens is over 1.7 million, with unique token types per speaker ranging from 28.5 thousand to 80.7 thousand.
Figure~\ref{fig:dist} presents the histograms of the distributions of sentence duration and length (in words) for each speaker in KazakhTTS2.
For all speakers, the majority of sentence durations are between 3 and 6 seconds. 
The majority of sentence lengths are between 11 and 15 words for female speakers, and between 6 and 10 words for male speakers.
%The duration of the audio utterances are between X and Y seconds, with the majority of the utterances having duration of X-Y seconds.
%The sentence lengths are between X and Y words, the majority of the sentences having Z words.

\begin{table*}[th]
    \begin{center}
        \small
        \renewcommand\arraystretch{1.1}
        \setlength{\tabcolsep}{2.07mm}
        \begin{tabularx}{\linewidth}{c|c|c|cccc|ccccc}
            \toprule
            \multirow{2}{*}{\textbf{Speaker ID}}& \multirow{2}{*}{\textbf{Source}}  & \multirow{2}{*}{\textbf{\# Segments}} & \multicolumn{4}{c|}{\textbf{Segment duration}}                &
            \multicolumn{5}{c}{\textbf{Tokens}} \\ \cline{4-12}
                                                &                                   &                                       & \textbf{Total}& \textbf{Mean} & \textbf{Min}  & \textbf{Max}  &
            \textbf{Total}  & \textbf{Mean} & \textbf{Min}  & \textbf{Max}  & \textbf{Unique}\\
            \midrule
            F1  & \ \ News$^*$  & 17,426    & 36.1 h    & 7.5 s & 1.0 s & 24.2 s    & 245.4k    & 14.1  & 2 & 42    & 34.3k \\ \hline
            F2  & News  & 12,921    & 25.7 h    & 7.2 s & 0.8 s & 22.0 s    & 177.9k    & 13.8  & 1 & 42    & 28.5k \\ \hline
            F3  & News  & 23,696    & 48.5 h    & 7.4 s & 0.7 s & 21.4 s    & 331.3k    & 14.0  & 1 & 48    & 43.0k \\ \hline
            \multirow{4}{*}{M1}     & \ \ News$^*$  & 24,608    & 57.0 h    & 8.3 s & 0.8 s & 55.9 s    & 319.6k    & 13.0  & 1 & 75    & 42.5k \\
                & Wiki  & 13,189    & 29.7 h    & 8.1 s & 0.9 s & 29.7 s    & 166.8k    & 12.6  & 1 & 47    & 33.6k \\
                & Book  & 11,453    & 16.7 h    & 5.3 s & 0.8 s & 21.1 s    & 107.6k    & 9.4   & 2 & 40    & 23.5k \\ \cline{2-12}
                & All   & 49,250    & 103.5 h   & 7.6 s & 0.8 s & 55.9 s    & 594.1k    & 12.1  & 1 & 75    & 80.7k \\\hline
            M2  & News  & 32,903    & 57.9 h    & 6.3 s & 0.7 s & 28.3 s    & 393.3k    & 12.0  & 1 & 60    & 53.3k \\
            \bottomrule
        \end{tabularx}
        
        \vspace{0.1cm}
        {\raggedright \textit{Note.} The news source data of speakers F1 and M1 are from KazakhTTS.\par}
        \caption{The KazakhTTS2 dataset specifications}
        \label{tab:data}
    \end{center}
\end{table*}

%##############################################################################################################################################################################################################
\section{Speech Synthesis Experiments}
In this section, we describe the experiments conducted to validate the utility of the KazakhTTS2 corpus.
We first describe the experimental setup, followed by our evaluation procedures and results.

\label{sec:exp}
\subsection{Experimental Setup}
We used the ESPnet-TTS toolkit~\cite{hayashi2020espnet} to build end-to-end TTS models based on the Tacotron~2 \cite{DBLP:conf/icassp/ShenPWSJYCZWRSA18} architecture.
Specifically, we followed the training recipe of LJ Speech~\cite{ljspeech17}.
All TTS models were trained using Tesla V100 GPUs running on NVIDIA DGX 2 machines.
The input for each model is a sequence  of characters consisting of 42 letters and 5 symbols (‘.’, ‘,’, ‘-’,‘?’, ‘!’), and the output is a sequence of acoustic features (80 dimensional log Mel-filter bank features).
To transform these acoustic features into the time-domain waveform samples, we employed WaveGAN~\cite{DBLP:conf/icassp/YamamotoSK20} vocoders. 

In the Tacotron 2 model, the encoder module was modeled as a single bidirectional LSTM layer with 512 units (256 units in each direction), and the decoder module was modeled as a stack of two unidirectional LSTM layers with 1,024 units.
The parameters were optimized using the Adam algorithm~\cite{DBLP:journals/corr/KingmaB14} with an initial learning rate of $10^{-3}$ for 200 epochs. 
To mitigate overfitting, we applied a dropout of 0.5.
A separate Tacotron 2 model was trained for each speaker (i.e., a single speaker model).
More details on the model specifications and training procedures are provided in our GitHub repository\textsuperscript{\ref{ft:github}}.

\subsection{Experimental Evaluation}
To assess the quality of the synthesized recordings, we performed a subjective evaluation using the MOS measure.
%We evaluated only the voices of the newly hired speakers (i.e., F2, F3, and M2), as the other two speakers had already been evaluated in the previous work\footnote{Note that, in this paper, we evaluated only news recordings, whereas the detailed evaluation and analyses of the recordings from other sources (i.e., Wikipedia and book) are left for future work.} \cite{mussakhojayeva21_interspeech}.
We evaluated only the voices developed using the newly collected data\footnote{For Speaker M1, we trained two separate models from scratch using Wikipedia and Book data.} (i.e., F2, F3, M1 Wikipedia and Book, and M2), as the other data had already been evaluated in the previous work \cite{mussakhojayeva21_interspeech}.
The evaluation procedure was similar to that of our previous work, except for the number of sentences selected as a test set.
Specifically, in this work, we selected 25 sentences of varying lengths from each speaker, whereas, in the previous work, 50 sentences per speaker were selected.
The reason for selecting a smaller number of sentences is based on our observation that raters become exhausted or bored after around 25 sentences and quit the evaluation session.
The evaluation sentences were not used to train the models.

The speakers were evaluated in separate sessions, and in each session we compared the ground truth (i.e., natural speech) recordings against the Tacotron 2 synthesized recordings.
The ground truth sentences were manually checked to ensure that the speaker read them well (i.e., without disfluencies, mispronunciations, or background noise).

%In particular, we instructed them to assess overall quality, to use headphones, and to sit in a quiet environment.
%(the full instruction in Kazakh and English are given in Appendix 1).
%The validation recordings were used among our lab members to decide the final E2E-TTS systems to use for evaluation.

Evaluation sessions were conducted using the instant messaging platform Telegram~\cite{telegram}, as it is difficult to find native Kazakh raters on other well-known platforms, such as Amazon Mechanical Turk~\cite{MTurk}.
We developed a separate evaluation Telegram bot for each speaker.
The bots first presented a welcome message with instructions and then started the evaluation process.
During the evaluation, the bots sent a sentence recording\footnote{Note that in Telegram, to send audio recordings, we had to convert them into MP3 format.} with the associated transcript to a rater and received the corresponding evaluation score. 
Recordings were rated using a five-point Likert scale: 5 for excellent, 4 for good, 3 for fair, 2 for poor, and 1 for bad.

The raters were instructed to assess the overall quality through headphones in a quiet environment\footnote{Due to the crowdsourced nature of the evaluation process, we cannot guarantee that all raters used headphones and sat in a quiet environment.}.
%We attracted listeners to participate in the evaluation sessions by advertising the project in social media, news, and open messaging communities on WhatsApp and Telegram.
They were allowed to listen to the recordings several times, but they were not allowed to alter the ratings once submitted.
Additionally, the Telegram bots kept track of the raters' ID, to prevent them from participating in the evaluation session more than once.
%As a result, the listeners could take a break to continue at a later time, and were prevented from participating in the evaluation session more than once.
%Additionally, they were allowed to take a break, and continue at a later time.

The evaluation recordings were presented in the same order and one at a time.
However, at each time step, the bots randomly decided which version of a recording to select (i.e., ground truth or synthesized).
%However, at each time step, the bots randomly decided from which system to pick a recording (i.e., ground truth or Tacotron 2).
As a result, each rater heard only one of the versions of a recording, and both systems (i.e., ground truth and Tacotron 2) were presented to all the raters.
%As a result, each examiner heard each recording once, but only one of its versions.
%Importantly, both systems (i.e., ground truch and Tacotron 2) were exposed to all the examiers.
Each recording was rated at least 24 times for all the three speakers.
The numbers of raters were 57, 61, 116, 89, and 53 for speakers F2, F3, M1 Wikipedia, M1 Book, and M2, respectively\footnote{In fact, the number of raters was higher, but we excluded the ratings of those who did not go through the session to the end, or whose ratings were suspicious (e.g., all scores are ``excellent" or all scores are ``bad".)}. 
%\footnote{In fact, the number of raters was higher, but we excluded ratings of those who did not go through the session to the end, or whose ratings for all recordings were suspiciously either ``excellent" or ``bad".}. 

At the end of the evaluation, the bots thanked the raters and invited them to fill in an optional questionnaire about their age, region (where a rater grew up and learned the Kazakh language), and gender.
The questionnaire results showed that the raters varied in gender and region, but not in age (most of them were under 20).
Specifically, the majority of raters were from the south and west of Kazakhstan, and females outnumbered males by a factor of 1.5.

\subsection{Experiment Results}
The subjective evaluation results are given in Table~\ref{tab:result}.
As expected, the ground truth recordings received higher MOS scores than the Tacotron 2 synthesized ones.
Nevertheless, all synthesized recordings except M1 Wikipedia scored above 4.0 on the MOS measure and were close to the ground truth (i.e., 8.7\%, 3.1\%, 18.1\%, 11.1\% and 5.2\% relative MOS reductions for speakers F2, F3, M1 Wikipedia, M1 Book, and M2, respectively).
These results demonstrate the utility of our KazakhTTS2 dataset for TTS applications. 
%Overall, the highest MOS score among the synthesized recordings was achieved by speaker F1, followed by M2 and then M1 News, F3.

Overall, the highest MOS score among the synthesized recordings was achieved by Speaker F1, and the lowest score was achieved by M1 Wikipedia.
Presumably, the reason for the poor performance of M1 Wikipedia is the wide variety of topics and the abundance of rare scientific terms (from chemistry, biology, information technology, etc.).
We believe that the performance of M1 Wikipedia can be improved by exploiting other data from Speaker M1.
For example, by pre-training a model on M1 News and Book data, followed by fine-tuning using M1 Wikipedia.
%According to the results, the best performance is achieved by the Ground truth, as expected, followed by the Tacotron 2, and then the Transformer system for both speakers.
%Importantly, the best performing models for both speakers achieved above 4 in the MOS measure and are not too far from the Ground truth, i.e., 4\% and 5\% relative reduction for the female and male speakers, respectively.

%The experimental results also indicate that the X hours of data is insufficient to build a decent E2E-TTS model, while there is not much difference between using \textbf{Y} and \textbf{Z} hours of data.
%Therefore, around \textbf{X} hours of annotated data is required to build a competent E2E-TTS model for the Kazakh language.

In addition, we conducted an objective evaluation in which we manually analyzed the synthesized evaluation set recordings. 
Specifically, we counted the various error types made by the Tacotron 2 systems built using the newly collected data.
The objective evaluation results are given in Table~\ref{tab:error}, which are consistent with the subjective evaluation, with Speaker M2 having the lowest number of errors, followed by F2 and M1 Book, and then F3 and M1 Wikipedia.
The most common error types among all speakers are mispronunciation, incomplete words, and word skipping.
%The mispronunciation errors are mostly due to incorrect stress, and the incomplete word errors mostly occur at the last word of a sentence, where the last letters of the word are trimmed.
%Interestingly, the total number of errors in the male speaker's recordings is considerably higher than the total number of errors in the female speaker's recordings.
%\textbf{This might be due to the poor hyper-parameter tuning in the male speaker's E2E-TTS model, as was mentioned in the previous section.}
%This might be one of the reasons for the lower MOS score achieved by the male speaker.
This analysis indicates that there is still room for improvement and future work should focus on eliminating these errors.

%We did not carry out an intensive hyper-parameter tuning for our Tacotron 2 model, since it is outside the scope of this work, therefore, we speculate that the quality of models can be further improved.
%%For example, our model training recipes are based on the LJ Speech, which is tuned for a dataset containing around 24 hours of audio, whereas our speakers' audio sizes are larger.
%We leave the exploration of the optimal hyper-parameter settings and detailed comparison of different TTS architectures for the Kazakh language as a future work.

%%The differences in performance can be attributed to many factors, such as quality and size of the data, neural network hyper-parameters, different raters and sample sentences, and so on.
%%We found the speaker F3's audios to be recorded under varying conditions with the inconsistent volume and reverberation. 

\begin{table}[t]
    \begin{center}
        \small
        \renewcommand\arraystretch{1.1}
        \setlength{\tabcolsep}{1.75mm}
        \begin{tabularx}{\linewidth}{cl|cc}
            \toprule
            %\multicolumn{2}{c|}{\textbf{Speaker ID}}& \textbf{Ground truth} & \textbf{Tacotron 2} \\
            \textbf{Speaker ID} & \textbf{Source}   & \textbf{Ground truth} & \textbf{Tacotron 2} \\
            \midrule
            F1                  & News$^*$          & 4.726 $\pm$ 0.037     & 4.535 $\pm$ 0.049 \\ \hline
            F2                  & News              & 4.449 $\pm$ 0.059     & 4.061 $\pm$ 0.070 \\ \hline
            F3                  & News              & 4.178 $\pm$ 0.073     & 4.049 $\pm$ 0.076 \\ \hline
            \multirow{3}{*}{M1} & News$^*$          & 4.360 $\pm$ 0.050     & 4.144 $\pm$ 0.063 \\
                                & Wiki              & 4.483 $\pm$ 0.040     & 3.673 $\pm$ 0.070 \\
                                & Book              & 4.564 $\pm$ 0.045     & 4.057 $\pm$ 0.068 \\ \hline
            M2                  & News              & 4.431 $\pm$ 0.062     & 4.200 $\pm$ 0.073 \\
            \bottomrule
        \end{tabularx}
        
        \vspace{0.1cm}
        {\raggedright \textit{Note.} The scores of speakers F1 and M1 News are from the previous work~\cite{mussakhojayeva21_interspeech}.\par}
        \caption{Mean opinion score (MOS) results with 95\% confidence intervals}
        \label{tab:result}
    \end{center}
\end{table}

\begin{table}[t]
  \setlength{\tabcolsep}{1.5mm}
  \small
  \centering
  \begin{tabular}{ l|c|c|c|c|c }
    \toprule
    \multirow{3}{*}{\textbf{Error types}}   & \multicolumn{5}{c}{\textbf{Speaker ID}} \\\cline{2-6}
                                            & \multirow{2}{*}{\textbf{F2}}  & \multirow{2}{*}{\textbf{F3}}  & \multicolumn{2}{c|}{\textbf{M1}}  & \multirow{2}{*}{\textbf{M2}} \\ \cline{4-5}
                                            &                               &                               & \textbf{Wiki} & \textbf{Book}     & \\
    \midrule
    \# repeated words                       & 0                             & 0                             & 0             & 0                 & 0 \\
    \# skipped words                        & 5                             & 6                             & 0             & 1                 & 1 \\
    \# mispronounced words                  & 2                             & 1                             & 11            & 5                 & 4 \\
    \# incomplete words                     & 2                             & 7                             & 0             & 3                 & 4 \\
    \# long pauses                          & 0                             & 0                             & 4             & 1                 & 0 \\
    \# nonverbal sounds                     & 1                             & 0                             & 0             & 0                 & 0 \\\hline
    \textbf{Total}                          & 10                            & 14                            & 15            & 10                & 9 \\
  \bottomrule
  \end{tabular}
  \caption{Manual analysis of error types made by Tacotron 2}
  \label{tab:error}
\end{table}

%##############################################################################################################################################################################################################
\section{Challenges and Future Work}
\label{sec:discuss}
The Kazakh language presents several challenges to the speech synthesis task.
The first one is code-switching, as the majority of Kazakh speakers are bilingual in Kazakh and Russian.
While the languages are not mixed in most formal situations (e.g., news, books, law, etc.), intrasentential code-switching often occurs in informal conversations.
Moreover, intra-word code-switching is also possible (e.g., Kazakh stem words with Russian suffixes or vice versa), which may further deteriorate TTS quality.

Additionally, Kazakh has a large number of loanwords from Russian, and these words usually retain the orthographic and phonological properties of the source language.
This has especially important consequences for TTS applications, as Russian differs from Kazakh in many aspects.
For example, in most Kazakh words, the stress is fixed on the final syllable, while in Russian, the stress can be on any syllable of a word~\cite{jouravlev2014stress}.
Furthermore, the spelling of Kazakh words closely matches their pronunciation, which is not the case with Russian words; for example, the letter ``\textit{o}'' is sometimes pronounced as \textipa{/a/}.
It is important to mention that due to globalization, the number of loanwords from other languages, especially English, is also increasing, which is likely to pose an additional challenge in the near future~\cite{DBLP:conf/specom/MussakhojayevaK21}.

Another challenge is that Kazakh is an agglutinative language, with a very large vocabulary and many characters per word.
It is also susceptible to morphophonemic changes arising during word formation.
One of the solutions would be to increase the size of the Kazakh speech corpus to cover more word formation variants.
We believe that overcoming these challenges for the Kazakh language will be an interesting direction for future research.

%##############################################################################################################################################################################################################
\section{Conclusion}\label{sec:conclude}
We have presented KazakhTTS2, a large-scale open-source Kazakh text-to-speech corpus, which further extends the previous work with more data, voices, and topics.
The corpus consists of five voices (three female and two male), with over 270 hours of high-quality transcribed data.
%The corpus is released under the CC BY 4.0 international license, which permits both academic and commercial use.
The corpus is publicly available, which permits both academic and commercial use.
We validated the corpus by means of crowdsourced subjective evaluation, where all voices synthesized using the Tacotron 2 model achieved an MOS of above 3.6, making it suitable for practical deployment.
To enable experiment reproducibility and facilitate future research, we shared our training recipes and pretrained models in our GitHub repository\textsuperscript{\ref{ft:github}}.
Although the corpus was designed with TTS application in mind, it can be used to complement other speech processing applications, such as speech recognition and speech translation.
We hope the TTS corpus construction and evaluation procedures described in this paper will contribute to the burgeoning field of Kazakh speech and language research and help advance the state-of-the-art for other low-resource languages of the Turkic family.

\section{Acknowledgements}
The authors would like to thank Aigerim Borambayeva, Almas Mirzakhmetov, Dias Bakhtiyarov, and Rustem Yeshpanov for their help in data collection, voice evaluation, and paper revision.
The authors would also like to thank the speakers for their recordings and the anonymous raters for their evaluations. %, and the anonymous reviewers for many helpful comments and suggestions.

%##############################################################################################################################################################################################################
%##############################################################################################################################################################################################################
%##############################################################################################################################################################################################################
%\subsection{Big tables}
%
%An example of a big table which extends beyond the column and will
%float in the next page.
%
% \begin{table*}[ht]
% \begin{center}
% \begin{tabular}{|l|l|}
%
%       \hline
%       Level&Tools\\
%       \hline\hline
%       Morphology & Pitrat Analyser\\
%       Syntax & LFG Analyser (C-Structure)\\
%       Semantics & LFG F-Structures + Sowa's Conceptual Graphs  \\
%       \hline
%
% \end{tabular}
% \caption{The caption of the big table}
% \end{center}
% \end{table*}
%
\iffalse
\section*{Appendix: How to Produce the \texttt{.pdf}}

In order to generate a PDF file out of the LaTeX file herein, when citing language resources, the following steps need to be performed:

\begin{itemize}
    \item{Compile the \texttt{.tex} file once}
    \item{Invoke \texttt{bibtex} on the eponymous \texttt{.aux} file}
 %   \item{Invoke \texttt{bibtex} on the \texttt{languageresources.aux} file}
    \item{Compile the \texttt{.tex} file twice}
\end{itemize}
\fi

% \nocite{*}
\section{Bibliographical References}\label{reference}
%\label{main:ref}

\bibliographystyle{lrec2022-bib}
\bibliography{main}

%\section{Language Resource References}
%\label{lr:ref}
%\bibliographystylelanguageresource{lrec2022-bib}
%\bibliographylanguageresource{languageresource}

\end{document}